\documentclass[aps,pra,preprint,showpacs,floatfix]{revtex4}
\usepackage{dcolumn}                 
\newcolumntype{w}[1]{D{.}{.}{#1}}

\newcommand*{\cent}[1]{\multicolumn{1}{c}{$#1$}}
\usepackage{times}
\usepackage{nicefrac}
\usepackage{amsmath}
\usepackage{amsfonts}
\usepackage{amssymb}
\usepackage{amsthm}

\begin{document}
\preprint{Version 2.0}
\title{Ground state of Li and Be$^+$ using explicitly correlated functions}

\author{Mariusz Puchalski}
\email[]{mpuchals@fuw.edu.pl}
\author{Dariusz K\c edziera}
\email[]{teodar@chem.uni.torun.pl}
\author{Krzysztof Pachucki}
\email[]{krp@fuw.edu.pl}

\affiliation{Institute of Theoretical Physics, University of Warsaw,
             Ho\.{z}a 69, 00-681 Warsaw, Poland}

\begin{abstract}
We compare the explicitly correlated Hylleraas and exponential
basis sets in the  evaluations of ground state of Li and Be$^+$.
Calculations with Hylleraas functions are
numerically stable and can be performed with the large number of 
basis functions. Our results for ground state energies  
$-7.478\,060\,323\,910\,10(32)$,  $-14.324\,763\,176\,790\,43(22)$
of Li and Be$^+$ correspondingly, are the most accurate to date.
When small basis set is considered, explicitly correlated exponential 
functions are much more effective. With only 128 functions we obtained
about $10^{-9}$ relative accuracy, but the severe numerical instabilities
make this basis costly in the evaluation.
\end{abstract}

\pacs{31.15.ac, 31.15.ve, 31.30.-i}
\maketitle

\section{Introduction}

In order to accurately calculate energy levels of light atomic systems, 
not only nonrelativistic energies, but also relativistic and QED 
corrections have to be obtained with the high precision. 
In the NRQED approach all corrections
are obtained perturbatively, in powers of the fine structure constant
$\alpha$. Each term of this expansion is expressed as 
the expectation value of some effective Hamiltonian with the
nonrelativistic wave function. Similarly, corrections due to the finite
nuclear mass and its size can all be included perturbatively.
This however requires the accurate representation of the nonrelativistic 
wave function.

The wave function of the ground and excited states
can be obtained on the base of the Ritz variational principle. 
The accuracy of the upper bound for energy mainly depends on the 
basis set of trial functions and
effectiveness of the optimization routine. There are not so
many possible choices of basis functions, knowing that 
electron correlations have to be accurately accounted for. 
The most serious problem in development of explicitly correlated methods 
is difficulty in accurate calculations of integrals appearing
in Hamiltonian matrix elements, and the complexity of these  integrals 
grows with the increasing number of correlated electrons.

The most often in use are correlated Gaussian functions which have been
applied so far to systems including up to six-electrons,
and the most accurate results in comparison to other methods,
have be obtained for Be atom \cite{beryllium1, beryllium2, beryllium3}. 
Relatively simple integrals and possible
generalization to systems with higher number of electrons is the
main advantage of Gaussian functions.
However, these functions have improper short-distance (Kato cusps) and
long-rage behavior. As a result, the convergence of the variational 
procedure is not very fast. Quality of the globally optimized trial functions, 
even in a few thousand basis set is often
insufficient for calculations of relativistic effects
beyond the leading order. In particular, we observe poor
convergence of matrix elements with singular operators i.e. Dirac $\delta$.

Until now, the most accurate nonrelativistic wave function for
lithium-like atomic systems were computed in Hylleraas basis by
King in \cite{king_lit}, by Yan and Drake in \cite{yan_lit} and by present
authors in \cite{lit_wave}. The Hylleraas function for 
the three-electron system is of the form
\begin{equation}
\phi(\vec r_1, \vec r_2, \vec r_3) =
r_{23}^{n_1}\,r_{31}^{n_2}\,r_{12}^{n_3}\,r_{1}^{n_4}\,r_{2}^{n_5}\,r_{3}^{n_6}
\,e^{-\alpha_1\,r_1-\alpha_2\,r_2-\alpha_3\,r_3}\,,
\label{01}
\end{equation}
with nonnegative integer values of $n_i$.
Although, algorithms for integrals with these functions are computationally
demanding, the correct long and short-range asymptotic and possibility to
use a large basis set of functions ($\sim$10000) with small number
of variational parameter ($\sim$15) allows one to achieve  high
accuracy. In a recent series of papers we formulated
the analytical method for calculations of Hylleraas integrals
with the help of recursion relations \cite{recursions}. 
In this work we tuned up the optimization routine
compared to our former work \cite{lit_wave}. As a result, we significantly
improved nonrelativistic energies as compared to the previously published
ones in \cite{lit_rel,yan_lit} and achieved about $10^{-14}$ precision. 

Even better precision can in principle be achieved with the
explicitly correlated exponential function.
In 1987 Fromm and Hill obtained the closed analytical formula
for the related four-particle integral
\begin{eqnarray}
g_0 &=& \int \frac{d^3 r_1}{4\,\pi}\,
\int \frac{d^3 r_2}{4\,\pi}\,
\int \frac{d^3 r_3}{4\,\pi}\, 
\frac{e^{-w_1\,r_1-w_2\,r_2-w_3\,r_3-
u_1\,r_{23}-u_2\,r_{13}-u_3\,r_{12}}}{r_{23}\,r_{31}\,r_{12}\,r_{1}\,r_{2}\,r_{3}}\,,
\label{02}
\end{eqnarray}
reducing the problem to the evaluation of multivalued
dilogarithmic functions of complex arguments \cite{fromm}. 
Their formula could be differentiated with
respect to the $w_a$ and $u_a$ to introduce pre-exponential powers
of the $r_a$ and $r_{ab}$, thus to generate the class of integrals 
needed for evaluation of Hamiltonian and overlap matrix elements. The
Fromm-Hill formula was modified later by Harris eliminating the necessity
of branch tracking on the complex plane
\cite{harris_slater}. Zotev and Rebane presented their
method for integrals with an extension to complex exponentials
\cite{zotev_pra}. They demonstrated fast convergence even in small  bases
and high potential of this method in variational calculations of
four-body systems \cite{rebane_ps2}. Recently, Guevara {\em et al}. \cite{turbiner}
have been able to optimize the correlated exponential function including
linear terms in inter-particle distances by the six-dimensional
numerical integration and obtained nonrelativistic energy with the relative
precision of about $10^{-3}$. 

Effectiveness of correlated exponential functions gives
opportunity to reduce significantly the size of the basis set
as compared to Gaussian and Hylleraas functions. However, the evaluation of 
corresponding integrals is the most time consuming part of the variational method. 
This fact suggests to use rather short basis with carefully optimized parameters. 
In this work these integrals are calculated as folows.
The master integral $g_0$ in Eq. (\ref{02}) is calculated using Harris formula
\cite{harris_slater}. Integrals with higher powers of inter-particle distances,
are obtained using recursion relations, which are derived from the
differential equation (\ref{20}).  As a demonstration of this method,
we performed numerical calculations of the nonrelativistic energy and
of Dirac-$\delta$ for the ground state of Li and Be$^+$. With 128 well
optimized correlated exponential functions with real parameters
we have obtained nonrelativistic energies with relative precision of about $10^{-9}$.
This precision is not impressive in comparison to the value
extrapolated from 13944 Hylleraas functions. However, the result
for lithium is comparable to six times bigger set of Hylleraas functions
or 1500 optimized Gaussians. The highly accurate wave function in a small basis
set gives a flexibility in development of numerical methods for
evaluation of more complicated integrals. It is  expected to be especially
valuable for evaluation of matrix elements of $m\,\alpha^6$ operators, which involves
integrals very difficult to deal with Hylleraas functions.

\section{Nonrelativistic wave function}
The ground state wave function $\Psi$ is represented as 
a linear combination of $\psi$, the antisymmetrized product 
of the spatial functions $\phi$ and the spin function $\chi$
\begin{eqnarray}
\psi &=& {\cal A}[\phi(\vec r_1,\vec r_2,\vec r_3)\,\chi]\,,
\label{05}\\
\chi &=&
\alpha(1)\,\beta(2)\,\alpha(3)-\beta(1)\,\alpha(2)\,\alpha(3)\,.
\label{06}
\end{eqnarray}
In the case of correlated exponential functions, 
$\phi(\vec r_1, \vec r_2, \vec r_3)$ is
\begin{equation}
\phi(\vec r_1, \vec r_2, \vec r_3) =
e^{-\alpha_1\,r_1-\alpha_2\,r_2-\alpha_3\,r_3-\beta_1\,r_{23}-\beta_2\,r_{13}-\beta_3\,r_{12}}\,,
\label{07}
\end{equation}
and we assume that $\alpha_i, \beta_i$ are real numbers.
These nonlinear parameters are subject of additional conditions.
Namely, when one of the electrons goes to infinity,
the wave function shall decay exponentially sufficiently fast, 
so for example $\alpha_1+ \beta_2+\beta_3  > \sqrt{2\,E_{\rm ion}}$, where
$E_{\rm ion}$ is the ionization energy. 

The expansion coefficients and nonlinear parameter are obtained by
minimization of energy with the Hamiltonian $H$
\begin{eqnarray}
H &=& T + V\,,
\label{08}\\
T = \sum_{a=1}^3\,\frac{\vec p_a^{\,2}}{2}, \qquad   V &=&
\sum_{a=1}^3\, -\frac{Z}{r_a} + \sum_{a>b=1}^3\,\frac{1}{r_{ab}}\,,
\label{09}
\end{eqnarray}
where $Z\,e$ is the nuclear charge and atomic units are used elsewhere.
After elimination of spin variables, the matrix element of $H$ can be
expressed as
\begin{eqnarray}
\langle\psi^L|H|\psi^R\rangle &=& \langle 2\,\phi^L(1,2,3)+
2\,\phi^L(2,1,3)- \phi^L(3,1,2)- \phi^L(2,3,1)- \phi^L(1,3,2)
\nonumber \\ &&
-\phi^L(3,2,1)| H\,|\phi^R(1,2,3)\rangle\,.
\label{10}
\end{eqnarray}
The individual matrix element $\langle \phi^L | H | \phi^R \rangle $ is
represented as a linear combination of 34 Slater integrals defined as
\begin{eqnarray}
g(n_1,n_2,n_3,n_4,n_5,n_6) &=& \int \frac{d^3 r_1}{4\,\pi}\,
                               \int \frac{d^3 r_2}{4\,\pi}\,
                               \int \frac{d^3 r_3}{4\,\pi}\,
                               e^{-w_1\,r_1-w_2\,r_2-w_3\,r_3-u_1\,r_{23}-u_2\,r_{13}-u_3\,r_{12}}
\nonumber \\ &&
r_{23}^{n_1-1}\,r_{31}^{n_2-1}\,r_{12}^{n_3-1}\,r_{1}^{n_4-1}\,r_{2}^{n_5-1}\,r_{3}^{n_6-1}\,,
\label{11}
\end{eqnarray}
where $n_i$ are nonnegative integers and $w_a = \alpha_a^L +
\alpha_a^R$, $u_a = \beta_a^L + \beta_a^R$. 
The number of necessary integrals for the matrix element of $H$ can be
significantly reduced. Rebane and Zotev \cite{rebane_ham} derived 
the formula which includes only seven integrals: the overlap integral
$\langle\phi^L|\phi^R\rangle$ and six Coulomb integrals
$\langle\phi^L|r^{-1}|\phi^R\rangle$, which we have found 
very useful. It reduces significantly the computational costs in most of
cases except for small $w_a, u_a$, where it becomes numerically
unstable. In this case  we use the numerically stable standard form
of the kinetic energy operator
obtained by direct differentiation of the left and the right
wave function over the electron coordinates.

\section{Calculation of Slater integrals}

\subsection{Integration by parts method}

The evaluation method of $g(n_1,n_2,n_3,n_4,n_5,n_6)$ 
in Eq. \eqref{11} is based on
the integration by parts identities, which are widely used
for the analytical calculation of Feynman diagrams~\cite{fdiag}. 
Let us consider the following integral  in the momentum space
\begin{eqnarray}
G(m_1,m_2,m_3;m_4,m_5,m_6) &=& \frac{1}{8\,\pi^6}\,\int d^3k_1\int
d^3k_2\int d^3k_3\, (k_1^2+u_1^2)^{-m_1}\,(k_2^2+u_2^2)^{-m_2} \nonumber\\
&&(k_3^2+u_3^2)^{-m_3}\,
(k_{32}^2+w_1^2)^{-m_4}\,(k_{13}^2+w_2^2)^{-m_5}\,(k_{21}^2+w_3^2)^{-m_6}
\label{12}
\end{eqnarray}
which is related to $g$ function by $g_0 \equiv g(0,0,0,0,0,0) =
G(1,1,1,1,1,1)$. There are 9 corresponding integration by parts identities 
\begin{eqnarray}
&&0 \equiv {\rm id}(i,j) = \int d^3k_1\int d^3k_2\int
d^3k_3\,\frac{\partial}{\partial\,{\vec k_j}}
 \Bigl[ \vec k_i\,(k_1^2+u_1^2)^{-m_1}
\nonumber \\ && (k_2^2+u_2^2)^{-m_2}\,(k_3^2+u_3^2)^{-m_3}
(k_{32}^2+w_1^2)^{-m_4}\,(k_{13}^2+w_2^2)^{-m_5}\,(k_{21}^2+w_3^2)^{-m_6}
\Bigr] , \label{13}
\end{eqnarray}
where $i,j=1,2,3$. The reduction of the scalar products from the
numerator leads to the relations between functions $G$ of different
arguments. These identities group naturally into three sets with
respect to $j$. For example for $j=3$ and $m_i=1$  we have 
the following system of three equations
\begin{eqnarray}
0 &=& G(0,1,1,1,2,1) - G(0,1,2,1,1,1) + G(1,0,1,2,1,1) -
G(1,1,0,1,2,1) \nonumber \\ && - G(1,1,0,2,1,1) + G(1,1,1,2,0,1) -
G(1,1,1,2,1,0) + G(1,1,2,1,0,1) \nonumber \\ && +
G(1,1,1,1,2,1)\,(-u_1^2 + u_3^2 - w_2^2) + G(1,1,2,1,1,1)\,(u_1^2 +
u_3^2 - w_2^2) \nonumber \\ && + G(1,1,1,2,1,1)\,(-u_2^2 + u_3^2 -
w_2^2 + w_3^2) . \nonumber \\
0 &=& G(0,1,1,1,2,1) + G(1,0,1,2,1,1) - G(1,0,2,1,1,1) -
G(1,1,0,1,2,1) \nonumber \\ && - G(1,1,0,2,1,1) + G(1,1,1,0,2,1) -
G(1,1,1,1,2,0) + G(1,1,2,0,1,1) \nonumber \\ && +
G(1,1,1,2,1,1)\,(-u_2^2 + u_3^2 - w_1^2) + G(1,1,2,1,1,1)\,(u_2^2 +
u_3^2 - w_1^2) \nonumber \\ &&
+ G(1,1,1,1,2,1)\,(-u_1^2 + u_3^2- w_1^2 + w_3^2) , \nonumber \\
0&=& G(0,1,1,1,2,1) + G(1,0,1,2,1,1) - G(1,1,0,1,2,1) -
G(1,1,0,2,1,1) \nonumber \\ && - G(1,1,1,1,1,1) +
2\,G(1,1,2,1,1,1)\,u_3^2 + G(1,1,1,2,1,1)\,(-u_2^2 + u_3^2 + w_1^2)
\nonumber \\ && + G(1,1,1,1,2,1)\,(-u_1^2 + u_3^2+ w_2^2) ,
\label{14}
\end{eqnarray}
Whenever $m_i=0$,  $G$ becomes a known two-electron integral $\Gamma$ as
defined in Appendix A. For example
\begin{eqnarray}
G(0,1,1;1,1,1) &=& 
\Gamma(-1,0,-1;w_2+w_3,w_1,u_2+u_3)\nonumber \\
&=& \frac{1}{2\,w_1}\,\biggl[
  {\rm Li}\biggl(1 - \frac{u_2 + u_3 + w_2 + w_3}{u_2 + u_3 + w_1}\biggr)
+ {\rm Li}\biggl(1 - \frac{u_2 + u_3 + w_2 + w_3}{w_1 + w_2 +
w_3}\biggr)\nonumber \\ &&
 + \frac{1}{2}\,\ln^2\biggl(\frac{w_1 + w_2 + w_3}{u_2 + u_3 + w_1}\biggr) + \frac{\pi^2}{6}\biggr].
\label{15}
\end{eqnarray}
We solve the system of equation (\ref{14}), for example against
$G(1,1,1;2,1,1)$, and obtain
\begin{equation}
\frac{1}{2}\,\frac{\partial\sigma}{\partial w_1}\,G(1,1,1;1,1,1)
-2\,w_1\,\sigma\,G(1,1,1;2,1,1) + P = 0\,, \label{16}
\end{equation}
where $\sigma$ is a polynomial
\begin{eqnarray}
\sigma &=& u_1^2\,u_2^2\,w_3^2  + u_2^2\,u_3^2\,w_1^2  +
u_1^2\,u_3^2\,w_2^2 + w_1^2\,w_2^2\,w_3^2
 + u_1^2\,w_1^2\,(u_1^2+w_1^2-u_2^2-u_3^2-w_2^2-w_3^2)
\nonumber \\ &&
 + u_2^2\,w_2^2\,(u_2^2+w_2^2-u_1^2-u_3^2-w_1^2-w_3^2)
 + u_3^2\,w_3^2\,(u_3^2+w_3^2-u_2^2-u_1^2-w_1^2-w_2^2)\,,\label{17}
\end{eqnarray}
and $P$ is a the sum of two-electron integrals $\Gamma$
\begin{eqnarray}
P &=& -u_1\,w_1\,[(u_1 + w_2)^2 - u_3^2] \,
\Gamma(0,0,-1;u_1+w_2,u_3,u_2+w_1) \nonumber \\ &&
 -u_1\,w_1\,[(u_1 + u_3)^2 - w_2^2]\,\Gamma(0,0,-1;u_1+u_3,w_2,w_1+w_3)
\nonumber \\ &&
 +[u_1^2\,w_1^2 + u_2^2\,w_2^2 - u_3^2\,w_3^2 + w_1\,w_2\,(u_1^2 + u_2^2 - w_3^2)]\,\Gamma(0,0,-1;w_1+w_2,w_3,u_1+u_2)
\nonumber \\ &&
 +[u_1^2\,w_1^2 - u_2^2\,w_2^2 + u_3^2\,w_3^2 +
w_1\,w_3\,(u_1^2 + u_3^2 -
w_2^2)]\,\Gamma(0,0,-1;w_1+w_3,w_2,u_1+u_3)
 \nonumber \\ &&
 -[u_2\,(u_2 + w_1)\,(u_1^2 + u_3^2 - w_2^2) - u_3^2\,(u_1^2 + u_2^2 - w_3^2)]\,\Gamma(0,0,-1;u_2+w_1,u_3,u_1+w_2)
\nonumber \\ &&
 -[u_3\,(u_3 + w_1)\,(u_1^2 + u_2^2 - w_3^2) - u_2^2\,(u_1^2 + u_3^2 - w_2^2)]\,\Gamma(0,0,-1;u_3+w_1,u_2,u_1+w_3)
\nonumber \\ &&
 + w_1\,[w_2\,(u_1^2 - u_2^2 + w_3^2) + w_3\,(u_1^2 + w_2^2 - u_3^2) ]\,\Gamma(0,0,-1;w_2+w_3,w_1,u_2+u_3)
\nonumber \\ &&
 +w_1\,[u_2\,(u_1^2 - w_2^2 + u_3^2) + u_3\,(u_1^2 + u_2^2 -
 w_3^2)]\,\Gamma(0,0,-1;u_2+u_3,w_1,w_2+w_3)\,.
 \label{18}
\end{eqnarray}
Since
\begin{equation}
G(1,1,1;2,1,1) = -\frac{1}{2\,w_1}\,\frac{\partial g_0}{\partial w_1}\,\label{19}
\end{equation}
Eq. \eqref{16} takes the form of a differential equation
\begin{equation}
\sigma\,\frac{\partial g_0}{\partial w_1} +
\frac{1}{2}\,\frac{\partial\sigma}{\partial w_1}\,g_0 + P = 0\,,
\label{20}
\end{equation}
or
\begin{equation}
\sqrt{\sigma}\frac{\partial}{\partial w_1}(\sqrt{\sigma}\,g_0) + P = 0\,. \label{21}
\end{equation}
Analogous differential equation with respect to other parameters
$w_i$ and $u_i$ can be obtain by appropriate permutation of
arguments, using the tetrahedral symmetry of the function $g_0$.
This differential equation has been previously derived in Ref. \cite{hh2}.

\subsection{Calculation of $g_0$}
$g_0$ was obtained in analytical form by Fromm and Hill in \cite{fromm}
in terms of combination of multivalued dilogarithmic function of complex arguments.
Their formula was later simplified by Harris \cite{harris_slater},
who was able to eliminate the ambiguity of choosing
the right branch of dilogarithmic function. 
In this work we use directly his formulae and allowed ourselves 
to verify its correctness. For this we used the solution of
the differential equation in terms of one-dimensional integral.
Namely, for $\sigma>0$ we find
\begin{equation}
g_0 = \frac{1}{\sqrt{\sigma}} \biggl(\int_{w_1}^{\infty} d w_1'
\frac{P(w_1')}{\sqrt{\sigma (w_1')}} + g_0\,\sqrt{\sigma}\bigr|_{w_1=\infty}\biggr),
\label{22}
\end{equation}
where
\begin{eqnarray}
g_0\,\sqrt{\sigma}\bigr|_{w_1=\infty} &=& \frac{{\rm sgn}(u_1)}{2}\,\biggl[
\frac{\pi^2}{6} + \frac{1}{2}\,\ln^2\biggl(\frac{u_1 + u_3 + w_2}{u_1 + u_2 +
  w_3}\biggr) +
{\rm Li}_2\biggl(1 - \frac{u_2 + u_3 + w_2 + w_3}{u_1 + u_3 + w_2}\biggr)
\nonumber \\ &&
+ {\rm Li}_2\biggl(1 - \frac{u_2 + u_3 + w_2 + w_3}{u_1 + u_2 + w_3}\biggr)\biggr].
\end{eqnarray}
The above integration over $w_1$ is performed numerically using adapted
Gaussian points for the logarithmic singularity at $w_1=\infty$, see 
Appendix of \cite{lit_rel}.

For $\sigma<0$ we find
\begin{equation}
g_0 = \frac{1}{\sqrt{-\sigma}}\,\int_{\tilde w_1}^{w_1} d w_1'
\frac{P(w_1')}{\sqrt{\sigma (w_1')}}, \label{24}
\end{equation}
where
\begin{equation}
\sigma|_{w_1=\tilde w_1}=0\,.
\end{equation}
This integral is performed numerically using Gauss-Legendre quadrature
in variable $t=\sqrt{w_1-\tilde w_1}$.
In the simplest case when $\sigma=0$, $g_0$ can be readily obtained from
Eq. (\ref{20})
\begin{equation}
g_0 = -2\,P\,\left(\frac{\partial\sigma}{\partial\,w_1}\right)^{-1}.
\end{equation}
In almost all the cases, we achieved 28 digits accuracy using quadruple precision
arithmetic with about 100 integration points. 

\subsection{Recurrence scheme}
Since the direct evaluation of $g(n_1,n_2,n_3,n_4,n_5,n_6)$ in Eq. \eqref{11}
is very time  consuming, it is desirable to derive 
recurrence relations permitting integrals of
larger index values to be expressed in terms of those with smaller
indices. From differential equation \eqref{21} we can deduce much more than
only integral representation for $g_0$. We notice that
\begin{equation}
g(n_1,n_2,n_3,n_4,n_5,n_6) = (-1)^{n_1 + \ldots + n_6}
\frac{\partial^{n_1}}{\partial w_1^{n_1}} \ldots
\frac{\partial^{n_6}}{\partial u_3^{n_6}}\;g_0.
\end{equation}
Analogously, we introduce $\sigma(n_1,n_2,n_3,n_4,n_5,n_6)$ and
$P(n_1,n_2,n_3,n_4,n_5,n_6)$ derived form $\sigma$ and
$P$ respectively. If $\sigma \neq 0$ then equation \eqref{20} takes the form
\begin{equation}
\frac{1}{2}\,\sigma(1,0,0,0,0,0)\,g(0,0,0,0,0,0) +
\sigma(0,0,0,0,0,0)\, g(1,0,0,0,0,0) = P(0,0,0,0,0,0)\,.
\label{simp}
\end{equation}
Clearly this algebraic equation can be used to obtain
$g(1,0,0,0,0,0)$ once $g(0,0,0,0,0,0)$ is evaluated from the direct Ref. \cite{harris_slater}
or integral (\ref{22},\ref{24}) formulae. 
Now, we differentiate equation \eqref{simp} $n_1-1$, $n_2$, $n_3$,
$n_4$, $n_5$, $n_6$ times over $w_1$,$w_2$, $w_3$, $u_1$, $u_2$,
$u_3$ respectively
\begin{equation}
\sum^{n_1...n_6}_{i_1...i_6=0}\binom{n_1}{i_1}_{1/2}..\binom{n_6}{i_6}_{1/2}
\sigma(n_1-i_1,...,n_6-i_6)\, g(i_1,..,i_6) =
P(n_1-1,n_2,n_3,n_4,n_5,n_6), \label{rec}
\end{equation}
where we introduced a Newton-like notation
\begin{equation} \binom{n}{0}_{1/2}=\frac{1}{2},\quad
\binom{n}{n}_{1/2}= 1,\quad \binom{n}{i}_{1/2} =
\binom{n-1}{i}_{1/2} + \binom{n-1}{i-1}_{1/2}.
\end{equation}
The above formula allows to express the integral $g(n_1,..,n_6)$ with
nonzero $n_1$ through $g$-integrals with smaller index values. 
The expression for $\sigma(n_1,n_2,n_3,n_4,n_5,n_6)$
can be explicitely generated as derivatives of the polynomial $\sigma$,
since they become zero for large values of indices $n_i$. 
$P$ has a simple structure in terms of two-electron integrals $\Gamma$
multiplied by a simple polynomial.
Derivatives of these polynomials can be calculated explicitly. 
For $\Gamma$ we use the recurrence scheme proposed by Korobov 
in \cite{kor_he}. 

Similar recurrence relations can be obtained from the differential
equation like that in Eq. (\ref{20}), but with respect to a different variable. 
We use them for the missing integrals with $n_1=0$
in the above $w_1$ scheme, thus completing the algorithm for all
$g$-integrals starting from the master one $g_0$. We use them also to check
the numerical stability of the recurrence scheme, as
$g(1,1,1,1,1,1)$ can be obtained from the differential equation
in any of these nonlinear parameters. As the result of this checking,
we found out, that these recursions become unstable for small values of
$\sigma$ in Eq. (\ref{17}) and as a remedy we used higher precision
arithmetics in this particular region.

Recently, Harris obtained a family of recurrence formulas which
enable construction of correlated exponential integrals with
arbitrary pre-exponential powers  of inter-particle distances
\cite{harris_rec}. In comparison to them, our recurrences are 
not equivalent. Harris's recurrences in the denominator
involve additional powers of $u_i$ and thus may become 
numerically unstable in the limit of small $u_i$.
This however, requires numerical verification.

\section{Optimization and results}

\subsection{Hylleraas basis set}

In Table I we present results obtained with Hylleraas
functions for ground states of Li and Be$^+$, as they are much more
accurate than previous ones in \cite{yan_lit, lit_wave}. 
In comparison to these former works, we 
used slightly different division into 5 sectors with 
its own set of nonlinear parameters
as proposed in Ref. \cite{yan_lit}, and enhanced the
optimization process by replacement of the minimization
routine with CG Polak-Ribberie \cite{polak} with modifications of the
line search algorithm \cite{more}. In Ref. \cite{lit_wave} we performed
optimization in quadruple precision arithmetics. Here we observe
that this precision is sufficient for determination of the nonrelativistic 
energy, but it is at the edge of numerical stability for analytical calculation
of gradients in a basis set corresponding to $\Omega\equiv$ max$(\sum_i n_i)=10$. 
Therefore, in this work we used sextuple precision arithmetics 
for the whole calculation. Obviously, optimization process 
in higher precision arithmetics takes more time, in this case it is about
5 times longer, but the accuracy is improved by at least an order of magnitude. 
The results presented in Table \ref{tab_hyll} are better 
than the former ones in 50 percent bigger basis set. 
Especially important is the numerical
result for maximum set of 13944 carefully optimized functions, 
as this guarantees good quality of extrapolation to $\infty$ and
estimation of an uncertainty.

\begin{table}[!hbt]
\caption{Ground state nonrelativistic energies for the ground state
of Li and Be$^+$ for various basis length with Hylleraas functions
with comparison to earlier results including correlated Gaussian functions.}
\label{table1}
\begin{ruledtabular}
\begin{tabular}{rll}
No. of terms & $E({\rm Li})$   & $E({\rm Be}^+)$ \\\hline
2625& -7.478\,060\,323\,570\,509  & -14.324\,763\,176\,517\,134   \\
4172& -7.478\,060\,323\,845\,785  & -14.324\,763\,176\,746\,865 \\
6412& -7.478\,060\,323\,898\,268  & -14.324\,763\,176\,783\,625  \\
9576& -7.478\,060\,323\,907\,743  & -14.324\,763\,176\,789\,144  \\
13944&-7.478\,060\,323\,909\,560  & -14.324\,763\,176\,790\,150  \\
$\infty$& -7.478\,060\,323\,910\,10(32) &  -14.324\,763\,176\,790\,43(22)
\\ \hline
9577$^{\rm a}$ & -7.478\,060\,323\,892\,4 &  -14.324\,763\,176\,766\,8  \\
$\infty^{\rm b}$ & -7.478\,060\,323\,906(8) &  -14.324\,763\,176\,784(11)  \\
$10000^{\rm c}$ & -7.478\,060\,323\,81 &    \\
$ 8000^{\rm d} $ &                      &  -14.324\,763\,176\,4 \\
$16764^{\rm e} $ &  -7.478\,060\,323\,451\,9 & \\
\end{tabular}
\begin{flushleft}
 a - Ref. \cite{yan_lit}, b - Ref. \cite{lit_rel}, c -
 Ref. \cite{stanke_nrli}, d - Ref. \cite{stanke_nrbei},
 e - Ref. \cite{sims}.
\end{flushleft}
\end{ruledtabular}
\label{tab_hyll}
\end{table}

\subsection{Correlated exponential basis set}

We optimized the correlated exponential basis set 
incrementally starting from 1 up to 128
functions as shown in Tables \ref{tab_sla2} and \ref{tab_sla3}. 
\begin{table}[!hbt]
\renewcommand{\arraystretch}{1.0}
\caption{Nonrelativistic energies and Dirac-$\delta$ expectation
values for the ground state of Li compared to results in Hylleraas
basis}
\begin{tabular}{rllll}
N & \cent{E({\rm Li})}& \cent{\delta E/E} & \cent{\delta (r_{a})} & \cent{\delta (r_{ab})}
\\ \hline \hline
1    &   -7.453\,907\,382   & $3.2\,10^{-3}$& 13.631\,327 & 0.614\,377 \\
2    &   -7.465\,318\,352   & $1.7\,10^{-3}$& 13.163\,649 & 0.617\,596 \\
4    &   -7.476\,009\,761   & $2.7\,10^{-4}$& 13.691\,905 & 0.586\,670 \\
8    &   -7.476\,936\,884   & $1.5\,10^{-4}$& 13.773\,519 & 0.576\,457 \\
16   &   -7.478\,052\,680   & $1.0\,10^{-6}$& 13.840\,924 & 0.545\,361 \\
32   &   -7.478\,059\,401   & $1.2\,10^{-7}$& 13.841\,641 & 0.544\,671 \\
64   &   -7.478\,060\,050   & $3.7\,10^{-8}$& 13.842\,162 & 0.544\,526 \\
96   &   -7.478\,060\,272   & $7.0\,10^{-9}$& 13.842\,641 & 0.544\,391 \\
128  &   -7.478\,060\,301   & $3.1\,10^{-9}$& 13.842\,618 & 0.544\,368 \\
\hline Hyll. $\infty$ & -7.478\,060\,323\,9 && 13.842\,610\,8 & 0.544\,324\,6 \\
\end{tabular}
\label{tab_sla2}
\end{table}
\begin{table}[!hbt]
\renewcommand{\arraystretch}{1.0}
\caption{Nonrelativistic energies and Dirac-$\delta$ expectation
values for the ground state of Be$^+$ compared to results in Hylleraas
basis}
\begin{tabular}{rllll}
N & \cent{E({\rm Be}^+)}& \cent{\delta E/E} & \cent{\delta (r_{a})} & \cent{\delta (r_{ab})}
\\
\hline  \hline
1    &  -14.269\,015\,274 & $3.9\,10^{-3}$& 34.584\,174 & 1.726\,084 \\
2    &  -14.319\,868\,303 & $3.4\,10^{-4}$& 34.818\,880 & 1.722\,376 \\
4    &  -14.324\,097\,014 & $4.7\,10^{-5}$& 35.163\,138 & 1.598\,315 \\
8    &  -14.324\,646\,319 & $8.2\,10^{-6}$& 35.082\,068 & 1.589\,484 \\
16   &  -14.324\,730\,041 & $2.3\,10^{-6}$& 35.118\,928 & 1.583\,949 \\
32   &  -14.324\,760\,432 & $1.9\,10^{-7}$& 35.109\,851 & 1.582\,886 \\
64   &  -14.324\,762\,726 & $3.1\,10^{-8}$& 35.102\,872 & 1.581\,131 \\
96   &  -14.324\,763\,106 & $4.9\,10^{-9}$& 35.105\,550 & 1.580\,752 \\
128  &  -14.324\,763\,141 & $2.5\,10^{-9}$& 35.105\,342 & 1.580\,583 \\
\hline Hyll. $\infty$ &  -14.324\,763\,176\,8 && 35.105\,055\,7 & 1.580\,538\,6
\end{tabular}
\label{tab_sla3}
\end{table}
At the starting point, the bigger basis was composed
of  previously optimized smaller basis and functions with randomly chosen 
nonlinear parameters under constraints resulting from interparticle
separation conditions. Due to the presence of many nonlinear parameters,
each function has its own set of 6 parameters, the optimization
process was divided into steps. In a single step nonlinear parameters 
of only one function were optimized using Powell method without gradient.
In one cycle all functions were optimized separately. For small basis
several cycles were needed to achieve convergence at the 9th digit 
after the decimal point,
and for larger set of functions number of cycles increases.  
Implementation is done in Fortran 95 in the quadruple
precision arithmetics. In the region of typical values of $w_a$ and
$u_a$, we observe very good numerical stability of recurrence relations.
However, in some particular cases during the minimization
process, where $\sigma$ in Eq. (\ref{17}) becomes small and changes its sign,
the sextuple precision arithmetics was needed, as the recurrence relations
lose numerical precision. The region of small $\sigma$ is numerically 
unstable and we have not found yet an alternative way of evaluation of
$g(n_1,n_2,n_3,n_4,n_5,n_6)$ functions, by avoiding the presence 
of $\sigma$ in the denominator. This would be necessary for 
larger basis set and for states with the higher angular momentum. 
The quadruple precision arithmetics for the
maximum basis of 128 functions guarantees high quality
of the total wave function and the energy. We observe
by comparison with Hylleraas results, that the relative accuracy of 
about $10^{-9}$ is achieved for energies, and about 5-6 significant digits 
for wave functions as indicated by the Dirac $\delta$ expectation values.

\section{Summary}
We have performed accurate calculations of the ground state energy and the wave
function of Li and Be$^+$ using explicitly correlated Hylleraas and
exponential basis sets. Obtained results with Hylleraas basis
are the most accurate to date, due to the use of large
number of functions and efficient optimization.
Results with correlated exponential functions are much less accurate,
but they are the most efficient for the limited number of functions.
The relative accuracy of about $10^{-9}$ for the nonrelativistic energy of the
ground state of Li and Be$^+$ with only 128 functions confirms high
effectiveness of this basis. Compared to both the  Hylleraas and the Gaussians
functions, it allows to reduce significantly the size of basis set. 
Using the computational method based on recurrence relations,
we are able for the first time to perform optimization process with 
as much as 128 correlated exppnential functions and even more,
if numerical instabilities for small $\sigma$ are eliminated,
probably by a different type of recurrences.

Our primary motivation for developing explicitly correlated exponential 
basis set is the efficient representation of the wave function in a
small number of basis functions. We aim to apply them for numerical
calculations of expectation values of operators corresponding to higher order
relativistic and QED effects. They involve integrals with quadratic inverse
powers of at least two interparticle distances. That kind
of integrals are very complicated in the evaluation in the Hylleraas 
basis set and have not yet been worked out by the recursion method 
of the authors. However, there is a know algorithm by King \cite{king_lit}, 
but his method is much too slow for a large scale computation. In the
case of Slater integrals the problem would even much more complicated,
but we think, one shall be able to perform this class of integrals 
numerically. Equiped with the large and accurately optimized Hylleraas basis \cite{lit_fine},
and with the short and flexible correlated exponential basis functions, we are aiming to
determine $m\,\alpha^6$ and $m\,\alpha^7$ effects in the hyperfine
and fine structure of lithium-like systems.

\section*{Acknowledgments}
This work was supported by NIST through Precision Measurement Grant
PMG 60NANB7D6153, DK acknowledges additional support from Foundation 
for Polish Science through the program START.

\appendix
\renewcommand{\theequation}{\Alph{section}\arabic{equation}}
\renewcommand{\thesection}{\Alph{section}}

\section{Two-electron integrals}
The two-electron integral $\Gamma$ is defined by
\begin{eqnarray}
\Gamma(n_1,n_2,n_3,\alpha,\beta,\gamma)
&\equiv&\int\frac{d^3r_1}{4\,\pi}\,\int\frac{d^3r_2}{4\,\pi}\,
e^{-\alpha\,r_1-\beta\,r_2-\gamma\,r_{12}}\,r_1^{n_1-1}\,r_2^{n_2-1}\,r_{12}^{n_3-1}.
\end{eqnarray}
This integral takes very simple form when all $n_i=0$
\begin{equation}
\Gamma(0,0,0,\alpha,\beta,\gamma) =
\frac{1}{(\alpha+\beta)\,(\alpha+\gamma)\,(\beta+\gamma)}.
\end{equation}
The explicit form for $n_i>0$ can be obtained by differentiation
with respect to the corresponding nonlinear parameter,
the result for negative $n_i$ is obtained by an integration,
for example
\begin{equation}
\Gamma(0,0,-1,\alpha,\beta,\gamma)  =
\frac{1}{(\alpha-\beta)\,(\alpha+\beta)}
\ln\biggl(\frac{\gamma+\alpha}{\gamma+\beta}\biggr).
\end{equation}
For the actual evaluation of $\Gamma$ we use compact reccurence relations
from the work of Korobov \cite{kor_he}.


\begin{thebibliography}{99}

\bibitem{beryllium1} J. Komasa, Chem. Phys. Lett. {\bf 363}, 307 (2002).

\bibitem{beryllium2} K. Pachucki and J. Komasa, Phys. Rev. Lett. {\bf 92}, 213001 (2004).

\bibitem{beryllium3} M. Stanke, D. K\c edziera, S. Bubin, and L. Adamowicz
                     Phys. Rev. Lett. {\bf 99}, 043001 (2007).

\bibitem{king_lit} P. J. Pelzl, G. J. Smethells, and F. W. King, 
                   Phys. Rev. E {\bf 65}, 036707 (2002);
                   D. M. Feldmann, P. J. Pelzl and F. W. King,
                   J. Math. Phys. {\bf 39}, 6262 (1998).

\bibitem{yan_lit}  Z.-C. Yan and G.W.F. Drake, Phys. Rev. A {\bf 52}, 3711 (1995);
         Z.-C. Yan, M. Tambasco, and G.W.F. Drake, Phys. Rev. A {\bf 57}, 1652 (1998);
         Z.-C. Yan, W. N\"ortersh\"auser and G.W.F. Drake, Phys. Rev. Lett. {\bf 100}, 243002
         (2008).

\bibitem{lit_wave} M. Puchalski and K. Pachucki, Phys. Rev. A {\bf 73}, 022503 (2006).

\bibitem{recursions} K. Pachucki, M. Puchalski and E. Remiddi,
     Phys. Rev. A {\bf 70}, 032502 (2004).

\bibitem{lit_rel} M. Puchalski and K. Pachucki, Phys. Rev. A {\bf 78}, 052511 (2008).

\bibitem{fromm} D. M. Fromm and R.N. Hill, Phys. Rev. A {\bf 36}, 1013 (1987).

\bibitem{harris_slater} F. E. Harris, Phys. Rev. A {\bf 55}, 1820 (1997).

\bibitem{zotev_pra} V.S. Zotev and T.K. Rebane, Phys. Rev. A {\bf 65}, 062501 (2002).

\bibitem{rebane_ps2} T.K. Rebane, V.S. Zotev and O.N. Yusupov,  
 Zh. Eksp. Theor. Fiz. {\bf 110}, 55 (1996), [JETP {\bf 83}, 28 (1996)].

\bibitem{turbiner} N.L. Guevara, F.E. Harris, A. V. Turbiner, arXiv: 0901.3020 (2009).

\bibitem{rebane_ham} T.K. Rebane, V.S. Zotev, Opt. Spektrosk. {\bf 75}, 945
 (1993), [Opt. Spectrosc. {\bf 75}, 557 (1993)].

\bibitem{fdiag} F.V. Tkachov, Phys. Lett. {\bf B100}, 65 (1981);
  K.G. Chetyrkin and F.V. Tkachov, Nucl. Phys. B {\bf 192}, 159 (1981).

\bibitem{hh2} K. Pachucki, Phys. Rev. A.{\bf 80}, 032520 (2009).

\bibitem{kor_he} V. I. Korobov, Phys. Rev. A {\bf 66}, 024501 (2002).

\bibitem{harris_rec} F.E. Harris, Phys. Rev. A {\bf 79}, 032517 (2009).

\bibitem{polak} E. Polak, {\it Computational Methods in Optimization. A Unified
   Approach.}  (Academic Press, New York 1971).

\bibitem{more} J. J. Mor\'e, and D.J. Thuente, ACM Trans. Math. Software {\bf 20}, 286 (1994).

\bibitem{stanke_nrli} M. Stanke, J. Komasa, D. Kedziera, S. Bubin,
  L. Adamowicz, Phys. Rev. A {\bf 78}, 052507 (2008).

\bibitem{stanke_nrbei} M. Stanke, J. Komasa, D. Kedziera, S. Bubin, 
  L. Adamowicz, Phys. Rev. A {\bf 77}, 062509 (2008)

\bibitem{sims} J.S. Sims and S.A. Hagstrom, {\em unpublished}.

\bibitem{lit_fine} M. Puchalski and K. Pachucki, Phys. Rev. A {\bf 79}, 032510 (2009).

\end{thebibliography}
\end{document}